\documentclass{natureprintstyle2}
\usepackage{graphicx,amsmath}
\usepackage{astjnlabbrev-nature} 

\usepackage{amssymb}
\title{Crack in the cosmological paradigm}  

\author{Eleonora Di Valentino$^{1,2}$}
\begin{document}
\maketitle

\begin{affiliations}
 \item Institut d'Astrophysique de Paris (UMR7095: CNRS \& UPMC- Sorbonne Universities), F-75014, Paris, France; e-mail: valentin@iap.fr\\
 \item Sorbonne Universit\'es, Institut Lagrange de Paris (ILP), F-75014, Paris, France\\
\end{affiliations}

\begin{abstract}
A time-dependent dark energy component of the Universe may be able to explain tensions between local and primordial measurements of cosmological parameters, shaking current confidence in the concept of a cosmological ‘constant’.
\end{abstract}

The measurements of Cosmic Microwave Background (CMB) temperature and polarization anisotropies obtained by the Planck satellite \cite{planck2015} have provided strong evidence for the $\Lambda$ cold dark matter ($\Lambda$CDM) cosmological model of structure formation. The $\Lambda$CDM model is based on many assumptions with only six free parameters, which presents a risk of oversimplifying the physics that drives the evolution of our Universe.
The most debatable assumption made in the $\Lambda$CDM scenario states that the mysterious dark energy (DE) component that produces the current accelerated cosmic expansion can be completely parameterized by a constant-in-time energy-density term, the cosmological constant $\Lambda$. However, tensions are arising between Planck and other cosmological measurements, which justify the study of possible extensions to  $\Lambda$CDM~\cite{DiValentino:2016hlg}. Writing in \texttt{Nature Astronomy}, Gong-Bo Zhao and collaborators~\cite{Zhao:2017cud} offer a way to relieve these tensions by introducing an evolving DE.

The nature of $\Lambda$, which is actually in agreement with the Planck data, is one
of the most significant unsolved problems in fundamental physics we have today.
As $\Lambda$ is assumed not to change with time, while both matter and radiation — the
other components of the Universe — evolve rapidly, it follows that the recent appearance of $\Lambda$ in the standard cosmological model implies an extreme fine-tuning of initial conditions. This fine-tuning is known in cosmology as the coincidence problem. Although it is possible that some tensions between the different experiments may be due to measurement systematics, it is interesting to explore whether alternatives to a constant $\Lambda$ can explain these discrepancies. The current most statistically relevant and intriguing disagreement is the value of the Hubble constant. In fact, the value reported by Riess et al.~\cite{R16} of $H_0=73.24 \pm 1.74$ km/s/Mpc at $68 \%$ confidence level, derived from local luminosity distance measurements, lies beyond three standard deviations from the most recent Planck result of $H_0=66.93\pm0.62$ km/s/Mpc at $68 \%$ confidence level~\cite{plancknewtau}. After several years of improved analyses and datasets, the tension between the CMB and the direct constraints not only persists but is increasing with time~\cite{Freedman:2017yms}. Could the current tensions therefore be considered as a first hint for new physics beyond  $\Lambda$?

The local estimate of the Hubble constant is based on the combination of different geometric distance calibrations of Cepheids, which yield three independent constraints~\cite{R16} on $H_0$ that are totally consistent with each other. Moreover, both the recent
determinations of $H_0$, from the H0LiCOW~\cite{holicow} strong lensing survey $H_0=71.9_{-3.0}^{+2.4}$ km/s/Mpc and from the type Ia supernovae as near-infrared standard candles~\cite{Dhawan:2017ywl} ($H_0=72.8 \pm 1.6 (stat) \pm 2.7 (syst)$ km/s/Mpc), go towards the value obtained from the local luminosity measurements~\cite{R16}. Conversely, the constraints obtained from the CMB data~\cite{plancknewtau} are more precise but model-dependent:
by assuming a specific scenario, theory and data are compared using a Bayesian approach. This model dependency implies that the constraints on a certain parameter can be significantly different by imposing a different theoretical framework. Moreover, the CMB bounds are affected by the degeneracy between the parameters that can induce similar effects on the observables.

\begin{figure}
\centering
\includegraphics[width=0.95\columnwidth]{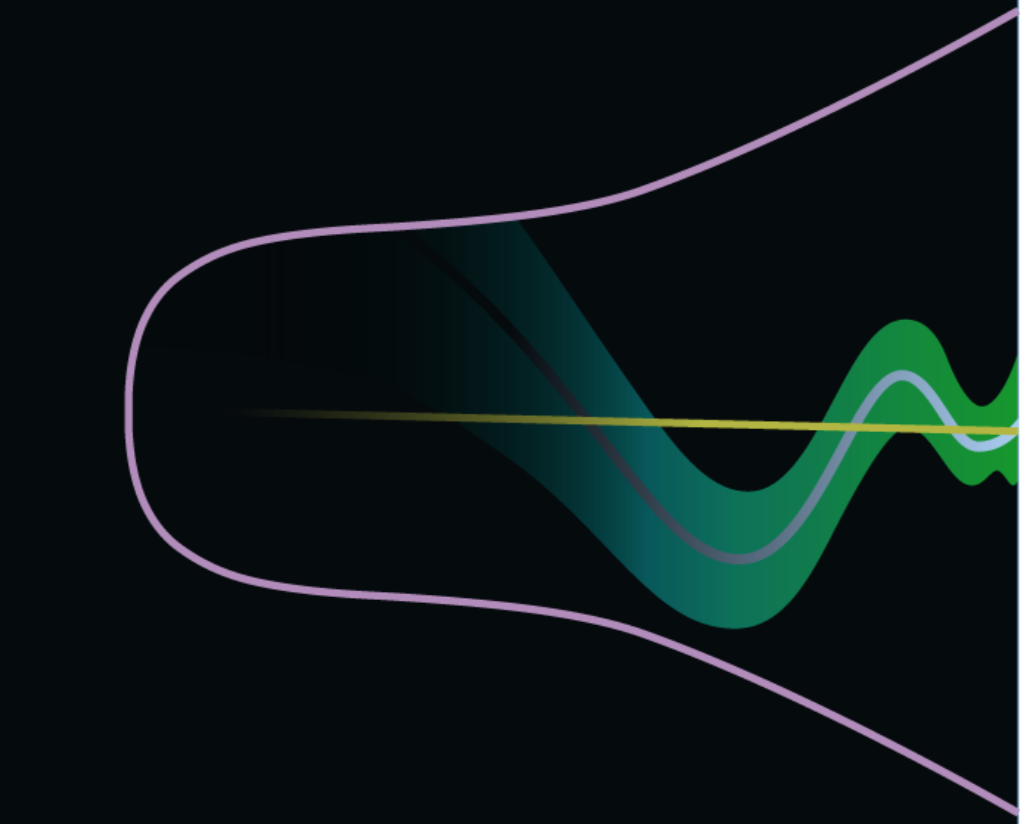} 
\caption{Time evolution of the dark energy equation of state. The cosmological ‘constant’ (illustrated by the straight yellow line) is introduced to explain the accelerated expansion of the Universe (shown as the expanding pink cone) due to the presence of dark energy. Zhao
et al.~\cite{Zhao:2017cud} instead suggest that the contribution of dark energy to this expansion is time-dependent (grey curve). The uncertainty of this time- dependency is also shown (green shaded area). Image credit: Gong-Bo Zhao, NAOC.}
\label{fig}
\end{figure}

In general, the DE evolution is expressed in terms of its equation of state $w$, defined as the ratio between the DE pressure $P_{DE}$ and energy density $\rho_{DE}$, where $w(z)=P_{DE}/\rho_{DE}$, which will be equal to $-1$ for $\Lambda$, but will be a function of redshift z in dynamical DE models. In the work by Zhao and colleagues, the evolution of $w$ is parametrized by varying its amplitude in different redshift bins from $z = 0$ up to the CMB last scattering surface at $z = 1100$. This model is the natural extension of the common alternative parameterizations to $\Lambda$ used in the literature. These include, usually, a model where $w$ is constant but differs from $-1$, or in which $w$ is a linear function of the scale factor (the Chevallier–Polarski– Linder parametrization). Both these models have already been suggested to solve the
$H_0$ tension~\cite{DiValentino:2016hlg,DiValentino:2017zyq}, thanks to the geometrical degeneracy that $w$ introduces with $H_0$, as both parameters modify the angular diameter distance at recombination. The importance of allowing an evolving
DE is that it can also overcome the coincidence problem with a dynamic solution that triggers a recent DE-dominated evolution of the Universe.
Zhao et al. analyse a consistent compilation of cosmological probes with the Kullback– Leibler (KL)~\cite{KL} divergence, which quantifies the degree of their disagreement with respect to an assumed cosmological model. By performing a Bayesian reconstruction of a time-dependent equation of state, $w(z)$, they find that these tensions are relieved by an evolving DE. The KL divergence indicates how much two probability density functions resemble each other, in a comparison of the overall concordance of datasets within a given model. The authors show that the $w(z)$CDM model results in an improved $\chi^2$ compared
to the $\Lambda$CDM model. The reconstructed DE equation of state that they obtained evolves with time (Fig.~\ref{fig}) crossing the $-1$ boundary, as in models with multiple scalar fields~\cite{Feng:2004ad} or in which the DE field mediates a new force between matter particles~\cite{Das:2005yj}. Moreover, a dynamical energy solves another potential conflict on the value of the matter density derived by the density fluctuations of baryons (baryon acoustic oscillations; BAO), which are traced by the large-scale structure of matter in the Universe.
Even if the dynamical DE model seems to provide a physical explanation for the $H_0$ disagreement, a model comparison based on Bayesian evidence is needed in order to understand which of the models is really favoured by the data. The authors found that, whereas the dynamical DE model is preferred at a $3.5\sigma$ significance level based on the improvement in the
fit of the data, the Bayesian evidence for the dynamical DE is insufficient to significantly favour it over$\Lambda$CDM with the currently available data. But Zhao et al. conclude that such dynamics could be decisively detected by the upcoming BAO measurements provided by the Dark Energy Survey Instrument (DESI) at higher redshifts. Clearly, future data from CMB experiments, such as the proposed Cosmic ORigin Explorer (CORE) satellite or ground-based telescopes such as Stage-4, and galaxy surveys, such as DESI and Euclid, will certainly clarify the issue, probably improving the determination of $w$ and $H_0$ by an order of magnitude and potentially resolving this cosmic conundrum.


\begin{thebibliography}{61}

\bibitem{planck2015} 
  P.~A.~R.~Ade {\it et al.} [Planck Collaboration],
  %``Planck 2015 results. XIII. Cosmological parameters,''
  arXiv:1502.01589 [astro-ph.CO]. 

%\cite{DiValentino:2016hlg}
\bibitem{DiValentino:2016hlg} 
  E.~Di Valentino, A.~Melchiorri and J.~Silk,
  %``Reconciling Planck with the local value of $H_0$ in extended parameter space,''
  Phys.\ Lett.\ B {\bf 761}, 242 (2016)
  doi:10.1016/j.physletb.2016.08.043
  [arXiv:1606.00634 [astro-ph.CO]].
  %%CITATION = doi:10.1016/j.physletb.2016.08.043;%%
  %51 citations counted in INSPIRE as of 19 Jul 2017

%\cite{Zhao:2017cud}
\bibitem{Zhao:2017cud} 
  G.~B.~Zhao {\it et al.},
  %``Dynamical dark energy in light of the latest observations,''
  Nat.\ Astron.\  {\bf 1}, 627 (2017)
  doi:10.1038/s41550-017-0216-z
  [arXiv:1701.08165 [astro-ph.CO]].
  %%CITATION = doi:10.1038/s41550-017-0216-z;%%
  %19 citations counted in INSPIRE as of 12 Sep 2017

\bibitem{R16}
  A.~G.~Riess {\it et al.},
  %``A 2.4% Determination of the Local Value of the Hubble Constant,''
  arXiv:1604.01424 [astro-ph.CO].
  %%CITATION = ARXIV:1604.01424;%%
  %1 citations counted in INSPIRE as of 23 Apr 2016

\bibitem{plancknewtau}
  N.~Aghanim {\it et al.} [Planck Collaboration],
  %``Planck intermediate results. XLVI. Reduction of large-scale systematic effects in HFI polarization maps and estimation of the reionization optical depth,''
  Astron.\ Astrophys.\  {\bf 596} (2016) A107
  doi:10.1051/0004-6361/201628890
  [arXiv:1605.02985 [astro-ph.CO]].
  %%CITATION = doi:10.1051/0004-6361/201628890;%%
  %71 citations counted in INSPIRE as of 12 Feb 2017

%\cite{Freedman:2017yms}
\bibitem{Freedman:2017yms} 
  W.~L.~Freedman,
  %``Cosmology at at Crossroads: Tension with the Hubble Constant,''
  Nat.\ Astron.\  {\bf 1}, 0169 (2017)
  [arXiv:1706.02739 [astro-ph.CO]].
  %%CITATION = ARXIV:1706.02739;%%
  %2 citations counted in INSPIRE as of 26 Jul 2017
  
 \bibitem{holicow}
 V.~Bonvin {\it et al.},
  %``H0LiCOW V. New COSMOGRAIL time delays of HE0435-1223: $H_0$ to 3.8% precision from strong lensing in a flat $\Lambda$CDM model,''
  doi:10.1093/mnras/stw3006
  arXiv:1607.01790 [astro-ph.CO].
  %%CITATION = doi:10.1093/mnras/stw3006;%%
  %13 citations counted in INSPIRE as of 16 Feb 2017 

%\cite{Dhawan:2017ywl}
\bibitem{Dhawan:2017ywl} 
  S.~Dhawan, S.~W.~Jha and B.~Leibundgut,
  %``Measuring the Hubble constant with Type Ia supernovae as near-infrared standard candles,''
  arXiv:1707.00715 [astro-ph.CO].
  %%CITATION = ARXIV:1707.00715;%%

%\cite{DiValentino:2017zyq}
\bibitem{DiValentino:2017zyq} 
  E.~Di Valentino, A.~Melchiorri, E.~V.~Linder and J.~Silk,
  %``Constraining Dark Energy Dynamics in Extended Parameter Space,''
  Phys.\ Rev.\ D {\bf 96}, no. 2, 023523 (2017)
  doi:10.1103/PhysRevD.96.023523
  [arXiv:1704.00762 [astro-ph.CO]].
  %%CITATION = doi:10.1103/PhysRevD.96.023523;%%
  %8 citations counted in INSPIRE as of 26 Jul 2017

%\cite{KL}
\bibitem{KL}
Kullback S., Leibler R. A.. 1951. Annals Math.Statist.,22,79

%\cite{Feng:2004ad}
\bibitem{Feng:2004ad} 
  B.~Feng, X.~L.~Wang and X.~M.~Zhang,
  %``Dark energy constraints from the cosmic age and supernova,''
  Phys.\ Lett.\ B {\bf 607}, 35 (2005)
  doi:10.1016/j.physletb.2004.12.071
  [astro-ph/0404224].
  %%CITATION = doi:10.1016/j.physletb.2004.12.071;%%
  %903 citations counted in INSPIRE as of 12 Sep 2017

%\cite{Das:2005yj}
\bibitem{Das:2005yj} 
  S.~Das, P.~S.~Corasaniti and J.~Khoury,
  %``Super-acceleration as signature of dark sector interaction,''
  Phys.\ Rev.\ D {\bf 73}, 083509 (2006)
  doi:10.1103/PhysRevD.73.083509
  [astro-ph/0510628].
  %%CITATION = doi:10.1103/PhysRevD.73.083509;%%
  %215 citations counted in INSPIRE as of 12 Sep 2017
  
\end{thebibliography}
\end{document}